# Lasing in 15 atm CO$_2$ cell optically pumped by a Fe:ZnSe laser

DANA TOVEY,[1,4] JEREMY PIGEON,[2] SERGEI TOCHITSKY,[1,5] GERHARD LOUWRENS,[1] ILAN BEN-ZVI,[2] DMITRY MARTYSHKIN,[3] VLADIMIR FEDOROV,[3] KRISHNA KARKI,[3] SERGEY MIROV,[3] AND CHAN JOSHI[1]

[1]*Department of Electrical Engineering, University of California at Los Angeles, California 90095, USA*
[2]*Department of Physics and Astronomy, Stony Brook University, Stony Brook, New York 11974, USA*
[3]*Department of Physics, University of Alabama at Birmingham, Birmingham, Alabama 35294, USA*
[4]*danatovey@ucla.edu*
[5]*sergei12@ucla.edu*

**Abstract:** 10 µm lasing is studied in a compact CO$_2$-He cell pressurized up to 15 atm when optically pumped by a ~50 mJ Fe:ZnSe laser tunable around 4.3 µm. The optimal pump wavelength and partial pressure of CO$_2$ for generating 10 µm pulses are found to be ~4.4 µm and 0.75 atm, respectively. Without cavity optimization, the optical-to-optical conversion efficiency reached ~10% at a total pressure of 7 atm. The gain lifetime is measured to be ~1 µs at pressures above 10 atm, indicating the feasibility of using high-pressure optically pumped CO$_2$ for the efficient amplification of picosecond 10 µm pulses.

## 1. Introduction

Picosecond and subpicosecond pulses with TW-level power in the long-wave infrared region (8-14 µm) are desirable for the study of high-field physics and nonlinear optics in this wavelength range [1-3]. Utilizing CO$_2$ laser systems is currently the most promising strategy towards the generation of such pulses, as the CO$_2$ molecule is capable of storing Joules of energy for 10 µm amplification and does not face the damage threshold limitations that inhibit optical parametric amplifiers from reaching high peak powers at these wavelengths. In addition, while the individual rovibrational transitions of the CO$_2$ gain spectrum have a relatively narrow bandwidth, these transitions can be broadened via collisions at high pressures to overlap and provide the THz bandwidth necessary for picosecond pulse amplification. CO$_2$ laser systems are typically pumped with an electric discharge, however, and the voltage required for this discharge scales linearly with pressure. As a result, at the high pressures (>10 atm) needed for a smooth gain bandwidth, it is extremely difficult to maintain a stable electric discharge in large volumes and repetition rates are limited.

Alternatively, the CO$_2$ active medium can be excited optically, removing this discharge pressure limitation. As shown in Figure 1(a), a pump source at ~4.3 µm can be used to directly populate the vibrational levels in the asymmetric stretching mode of the CO$_2$ molecule, which contains the upper laser level 001 for the 10 µm laser transitions of interest. The optical pumping of a CO$_2$ laser was historically first demonstrated using an incoherent CO-air flame to excite CO$_2$, producing ~1 mW of cw radiation at 10.6 µm [4]. Later, using a pulsed ~4.23 µm HBr laser as a pump source, lasing was observed in optically pumped CO$_2$ at 33 atm total pressure, but optical-optical conversion efficiency was not optimized or reported [5]. In addition, the cavity was constricted to only ~1 mm in length, reducing the round-trip time to combat the short gain lifetimes at such high pressures but also decreasing the single-pass gain in such a short CO$_2$ cell. The 101 level of the CO$_2$ molecule has also been pumped using a ~2.7 µm HF chemical laser [6,7], but this method was not pursued beyond the initial experimental demonstration, as the fast collisional decay of the 101 level simultaneously populates both the upper (001) and lower (100) laser levels for the 10 µm band. In general, the lack of available energetic pump sources operating at wavelengths that are efficiently absorbed by $CO_2$ has prevented the demonstration of high gain in a high-pressure optically pumped $CO_2$ amplifier.

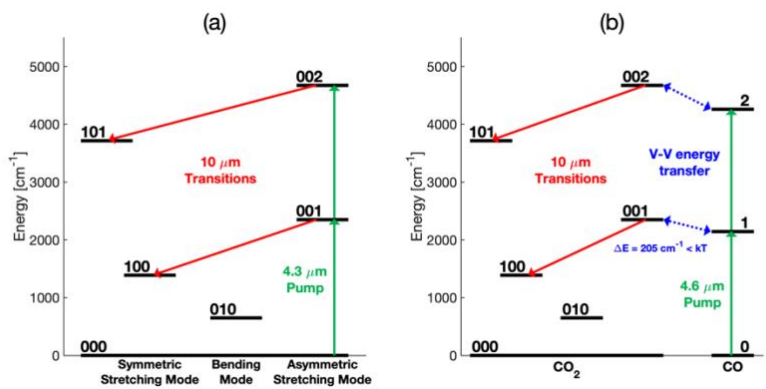

Figure 1. Optical pumping scheme for (a) direct excitation of $CO_2$ and (b) indirect excitation of $CO_2$ via CO. Green arrows indicate absorption of pump photons, while red arrows indicate laser transitions.

An alternative optical pumping scheme has also been explored, in which a partner molecule is excited by optical pumping and energy is then transferred to $CO_2$ via collisions. This method enables the use of pump wavelengths that do not overlap directly with $CO_2$ absorption lines and can also result in significantly longer gain lifetimes due to the slow exchange of vibrational energy. Figure 1(b) depicts such a scheme using CO as the collisional partner molecule to be optically excited. The small quantum defect between the first vibrational level of CO and the first level of the $CO_2$ asymmetric stretching mode allows energy to be efficiently transferred from CO to $CO_2$. Lasing was demonstrated in a CO-$CO_2$-He mix at a total pressure of 16 atm by pumping CO at 4.8 µm using the second harmonic of a 9.6 µm TEA $CO_2$ laser [8]. Less than 1 mJ of 10 µm light was generated, however, as this method faces a similar lack of energetic pump sources.

Recently, progress has been made on solid-state lasers using iron-doped zinc chalcogenides to provide tunable sources around 3–5 µm that have the potential to be scaled to generate pulses with Joule-class energies [9]. Previously, we have demonstrated high conversion efficiencies of ~30% and high gain coefficients in a low pressure (≤1 atm) $CO_2$ active medium pumped by ~2 mJ pulses generated by a Fe:ZnSe laser [10]. In this paper, we study a $CO_2$ cell optically pumped by a 50 mJ, tunable Fe:ZnSe MOPA system to determine the feasibility of generating short 10 µm pulses in a high-pressure optically pumped $CO_2$ active medium. Two pumping schemes were studied experimentally, one in which $CO_2$ was directly excited at ~4.3 µm (see Figure 1(a)), and one in which $CO_2$ was indirectly excited via pumping CO at ~4.6 µm (see Figure 1(b)). In the direct $CO_2$ excitation scheme, lasing and gain were measured at total pressures up to 15 atm. The effect of absorption coefficient on lasing performance was determined by varying the pump wavelength, and small-signal gain measurements were used to experimentally measure gain lifetime as a function of pressure. Finally, the rate of dissipation of the heat or pressure waves generated by the pump pulse inside the cell was measured and used to estimate the maximum repetition rate of an optically pumped $CO_2$ system.

## 2. Experimental techniques

The experimental setup used to study lasing in an optically pumped $CO_2$ cell is shown in Figure 2(a). The green arrows illustrate the path of the ~200 ns pump pulses generated by the Fe:ZnSe MOPA system, which contain ≤60 mJ of energy and can be tuned over a wavelength range of ~3.8–5.0 µm. This system consists of a master oscillator and three stages of amplification, each of which are pumped by a mechanically Q-switched, 2.94 µm Er:YAG pump laser. Further details on the Fe:ZnSe laser can be found elsewhere [11]. The output pulse was focused to a spot size of ~3.7 mm, corresponding to an intensity of ≤2.8 MW/cm$^2$ and a fluence of ≤500 mJ/cm$^2$. A NaCl wedge angled at 45 degrees was used to sample ~1% of the pump pulse to diagnose the input pump energy prior to reaching the $CO_2$ cell as shown. For all experiments discussed here, the entire experimental area was purged with $N_2$ gas to eliminate any absorption of the pump pulse by $CO_2$ in the ambient air.

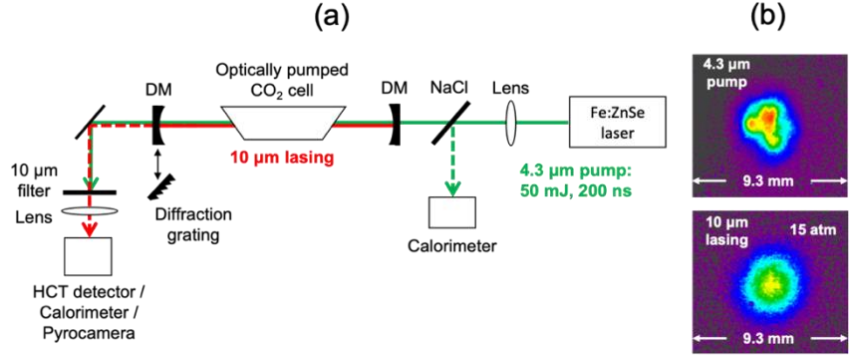

Figure 2. (a) Experimental setup of the optically pumped $CO_2$ laser. DM – dichroic mirror, the left DM can be replaced by a diffraction grating. Spatial beam profiles of the 4.3 µm pump laser (top) and the 10 µm optically pumped $CO_2$ laser (bottom).

The optically pumped $CO_2$ laser cavity was formed by two curved dichroic mirrors surrounding a 6 cm long high-pressure gas cell sealed with Brewster-angled NaCl windows. The dichroic mirrors are 99.5% reflective at ~10.6 µm and 99% transmissive at ~4.3 µm, and have a radius of curvature of 70 cm. The cavity length was ~50 cm. The 10.6 µm laser radiation decoupled through the downstream (leftmost in Figure 2(a)) dichroic mirror was measured as shown. A 10 µm filter was used to eliminate any residual 4.3 µm pump radiation. Typical spatial beam profiles of the 4.3 µm pump pulse and the 10.6 µm laser pulse at high pressure are shown in Figure 2(b). As indicated in Figure 2(a), one of the dichroic mirrors could be replaced with a diffraction grating to study the effect of varying the cavity Q factor. For this configuration, the zeroth order reflection off of the grating was used to decouple radiation out of the cavity. Two separate diffraction gratings were used in this study, with measured reflectivities of ~71% and ~97.7% at 10.6 µm and groove densities of 135 gr/mm and 150 gr/mm, respectively.

A similar setup was used to measure small-signal gain and gain lifetime in optically pumped $CO_2$. For this purpose, the left dichroic mirror was removed, and a commercial low-pressure discharge-pumped pulsed 10.6 µm $CO_2$ laser was used to probe the active medium. Typical probe pulses contained a few millijoules of energy and consisted of a short (~100 ns) spike followed by a long (>20 µs) tail. The probe beam was aligned through the cell and reflected off the right dichroic mirror at a slight angle, measuring small-signal gain in a V-shaped double-pass scheme to prevent cross-interaction between the pump and probe lasers.

To measure the effect of heat or pressure waves generated in the cell by the almost completely absorbed pump pulse, the active medium was also probed by a visible laser. For this measurement, a 632.8 nm He-Ne laser was used to determine the rate at which perturbations caused by the heat waves dissipate. A ZnSe beam splitter was used to couple together the probe and pump beams, allowing the He-Ne laser to probe the cell in a linear double-pass scheme before being measured on a Si photodiode detector. To avoid damaging the photodiode, the residual 4.3 µm pump pulse was filtered in a glass window. By observing a reduction of the probe signal after the arrival of the Fe:ZnSe pump pulse caused by deflection of the probe beam and the subsequent recovery of the signal, a physical dissipation rate was estimated.

## 3. Results and discussion

### 3.1 Lasing in direct $CO_2$ pumping scheme

To study the direct $CO_2$ pumping scheme, the cell was filled with various mixtures of $CO_2$ and Helium gas and pumped at ~4.3 µm to directly excite the $CO_2$ asymmetric stretching mode as detailed in Figure 1(a). Figure 3(a) shows the normalized absorption spectrum at ~4.3 µm for a mixture of 0.75 atm $CO_2$ and 9.25 atm Helium, which is close to the optimal ratio as will be discussed below. The effect of changing absorption strength for a given mixture was studied by varying the wavelength of the continuously tunable Fe:ZnSe pump laser along this $CO_2$ absorption spectrum. The red markers in Figure 3(a) indicate the five pump wavelengths studied in this experiment, ordered from maximum (~3.5 $cm^{-1}$) to minimum (~0.2 $cm^{-1}$) absorption. Wavelengths were chosen to study a range of absorption from the peak to the wing of the distribution, and 4.256 µm (marker 2 in Figure 3(a)) was chosen specifically because it corresponds to the fourth sub-harmonic of the wavelength of a Nd:YAG laser (1064 nm), a possible alternative pump source [12,13].

Figure 3(b) displays the maximum total pressure at which lasing was observed as a function of measured absorption coefficient for a cell filled with different partial pressures of $CO_2$ and ballast He. The numbers next to each marker indicate the corresponding pump wavelength as shown in Figure 3(a), and the dashed lines indicate a calculated absorption coefficient that would correspond to 50% and 99% of the pump energy being absorbed over the 6 cm cell length. Note that for each of the markers to the right of the second dashed line, the entire pump pulse energy was absorbed.

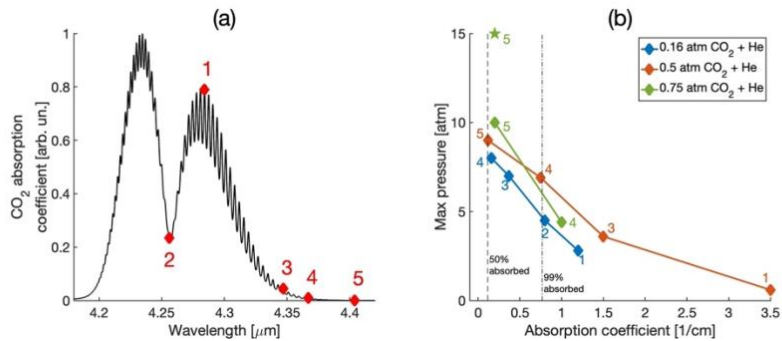

Figure 3. (a) Normalized absorption in 0.75 atm $CO_2$ and 9.25 atm He. Red markers indicate pump wavelengths used in experiment. (b) Maximum lasing pressure vs estimated experimental absorption coefficient. Numbers correspond to pump wavelengths indicated in (a). The green star indicates maximum lasing pressure after increasing the pump intensity by a factor of ~2 to ~5.3 MW/$cm^2$.

It can be seen that tuning the pump wavelength far from the peak of the absorption spectrum was necessary to achieve lasing at high pressures. This is attributed to the fact that in high absorption mixtures, nearly all of the pump energy was absorbed over a short distance, reducing the gain length of the $CO_2$ active medium. Increasing the partial pressure of $CO_2$ allowed for increased gain coefficients, but this also increased absorption and thus reduced the gain length. Note that lasing was not achieved at pressures above 5 atm when the cell was pumped at 4.256 µm. The relatively high absorption coefficient here dictates that, if using an alternate pump source limited to this wavelength, efficient short pulse amplification will require cell length optimization or utilizing a transverse pump configuration.

For high-pressure lasing in a 6 cm cell, the optimal partial pressure of $CO_2$ was found to be 0.75 atm and the optimal pump wavelength was found to be 4.40 µm. Further focusing of the pump beam to increase the intensity by a factor of ~2 to ~5.3 MW/cm$^2$ allowed for the achievement of lasing at a total pressure of 15 atm, as indicated by the green star in Figure 3(b). Note that the maximum pressure in this experiment was ultimately limited by the gas handling system, and higher pressures can be reached with further optimization. In addition to advantageously increasing the absorption length in the cell, pumping this far from the peak of absorption eliminates the need for purging the experimental area with $N_2$, as $CO_2$ absorption in the ambient air at this wavelength is negligible. It should be noted that the experimentally measured absorption using high intensity, 4.40 µm radiation was significantly higher than that theoretically predicted using HITRAN molecular constants [14]. This can be attributed to the fact that, within the cell, there may be significant absorption on the 001-002 and 002-003 absorption bands, as all vibrational levels in the asymmetric mode are populated and the absorption transitions of these sequence bands are shifted to longer wavelengths. This may also indicate a non-Boltzmann distribution of population among rovibrational energy levels for highly excited $CO_2$ molecules and requires further study.

The optical-to-optical conversion efficiency was also measured using several different output couplers with this same optimal $CO_2$ pressure and pump wavelength. Figure 4(a) shows the conversion efficiency as a function of total pressure for cavities formed by a second dichroic mirror, a 97.7% reflective diffraction grating, and a 71% reflective diffraction grating as indicated. It should be noted that flat reflective diffraction gratings were used, while the dichroic mirror had a radius of curvature of 70 cm, which likely helped reduce diffraction losses and therefore the lasing threshold. Figure 4(b) shows the same set of data, here presenting the conversion efficiency as a function of cavity Q factor. It was found that the highest reflectivity output coupler provided the best performance both in terms of maximum achievable lasing pressure and in terms of highest conversion efficiency.

Figure 4(c) shows measurements of 10.6 µm lasing energy as a function of absorbed 4.40 µm energy at a total pressure of 7 atm for the cavity utilizing a second dichroic mirror as an output coupler. Note that for this measurement, a range of absorbed energy was covered not by tuning wavelength but by attenuating the pump pulse energy by a factor of 2, and small variations in the data reflect the day-to-day change of the output of the Fe:ZnSe pump laser. There is no sign of saturation with increased pump energy, indicating the potential for scaling 10 µm output to higher energies.

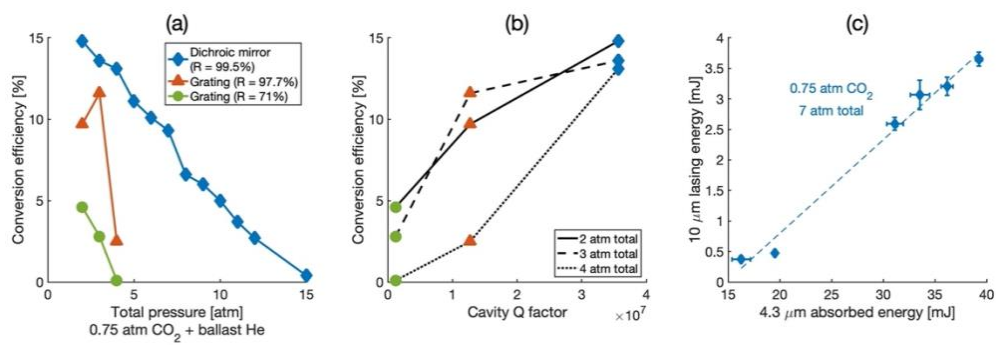

Figure 4. (a) Optical-optical conversion efficiency of the optically pumped $CO_2$ laser as a function of pressure using three different output couplers. (b) Conversion efficiency as a function of cavity Q factor at three different pressures. (c) 10 µm lasing energy generated vs. 4.40 µm pump energy absorbed at 7 atm total pressure.

*3.2 Small-signal gain in direct $CO_2$ pumping scheme*

Small-signal gain measurements were also performed to study the gain lifetime of an optically pumped $CO_2$ active medium. While in the past, relaxation rates have been measured in $CO_2$ systems pumped at ~4.3 µm [15,16], the extreme excitation of the $CO_2$ asymmetric stretching mode that is presented in this study may lead to different relaxation rates and has not been analyzed. To study this, small-signal gain and gain lifetime were measured by probing an optically pumped $CO_2$ system as described previously in section 2.

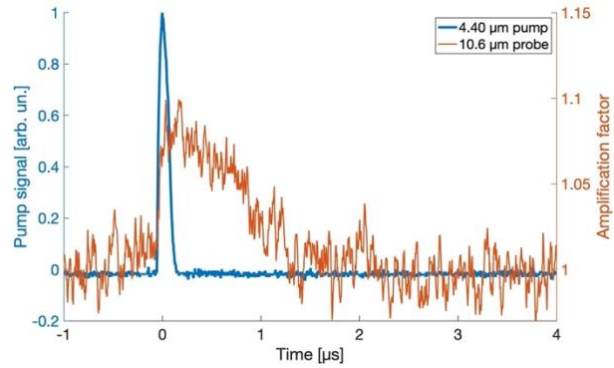

Figure 5. The 4.40 µm pump pulse (blue) and the amplification factor of the 10.6 µm probe pulse (red) as a function of time for a gas mix of 0.75 atm $CO_2$ and 10.25 atm He.

Figure 5 shows a typical measurement of the 4.40 µm pump pulse (in blue) and the total amplification factor of the 10.6 µm probe pulse (in red) as a function of time for an example gas mix of 0.75 atm $CO_2$ and 10.25 atm He. It can be seen that amplification reaches a maximum after the termination of the pump pulse, indicating the importance of repopulating the upper laser level by intramode collisional energy exchange to reach the maximum gain coefficient. Figure 6(a) shows the peak of this total amplification factor as a function of pressure. The blue and red curves correspond to measurements made in pure $CO_2$ and in 0.75 atm $CO_2$ mixed with ballast He, respectively. Fitting an exponential curve to the decay of these measured gain signals over time resulted in the measured gain lifetimes displayed in Figure 6(b). The dashed lines in Figure 6(b) correspond to values calculated using the experimentally determined relaxation rates for $CO_2$-$CO_2$ and $CO_2$-He collisions measured by Inoue et. al. and Lepoutre et. al. [15,16]. From the data in Figure 6(b), it is clear that our gain lifetimes agree well with these previously published values, indicating that for our pumping conditions, the physical processes occurring in an optically pumped $CO_2$ active medium do not differ significantly from those published in literature for a lower degree of excitation. Importantly, a long gain lifetime of ~1 µs at high pressures ≥10 atm opens the opportunity for building both regenerative and multi-pass amplifiers suitable for amplifying short seed pulses. Thus, in the direct optical pumping scheme using 4.4 µm Fe:ZnSe lasers, a high conversion efficiency of ~10% could be reached when the stored energy is extracted in a short pulse.

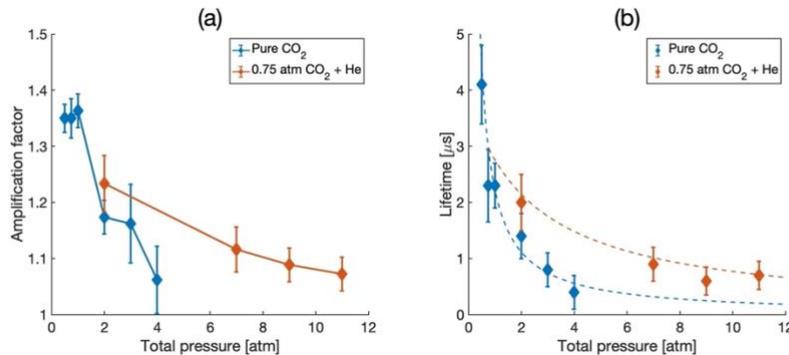

Figure 6. (a) Total amplification of a 10.6 µm probe in the 6 cm optically pumped $CO_2$ cell and (b) gain lifetime as a function of total pressure.

### 3.3 Lasing in indirect CO/CO$_2$ pumping scheme

Lasing was also studied in CO-CO$_2$-He mixtures utilizing the indirect pumping scheme, in which CO molecules are optically excited and energy is collisionally transferred to the CO$_2$ asymmetric stretching mode and thus the upper laser level (see Figure 1(b)). This scheme is motivated by the potential of further increasing the gain lifetime at high pressures due to the slow exchange of vibrational energy in a dual molecular system where the absorbing two-atomic molecule has a long relaxation time [8]. This scheme may also be used to avoid deleterious nonlinear refraction that can occur in CO$_2$ [17], which has a dipole moment three times larger than that of CO.

The same setup detailed in Figure 2(a) was used for these measurements, with the Fe:ZnSe pump laser wavelength now tuned to ~4.6 µm to coincide with absorption in the 0-1 vibrational band of CO. Figure 7(a) shows the normalized absorption spectrum of 1 atm CO, and the red markers indicate the pump wavelengths used in this experiment. Figure 7(b) shows the maximum total pressure at which 10.6 µm lasing was analyzed for two different gas mixtures as shown. The two dashed lines again indicate the absorption coefficients at which 50% and 99% of the pump pulse energy are absorbed over a length of 6 cm, and all pump energy was absorbed for the rightmost two data points. Note that lasing at pressures above 4 atm was not achieved in either the 9:1 CO$_2$:CO or the 10:3:27 CO$_2$:CO:He mixes, and varying the absorption by tuning pump wavelength did not have an observable positive effect on the maximum lasing pressure. While we have observed a gain lifetime that is several times longer than that for direct pumping, the gain values and 10 µm output stability were clearly inferior to those measured in the case of 4.4 µm pumping of the CO$_2$ medium.

For both mixtures, an audible noise corresponding to heat or pressure waves inside the cell could be heard upon the arrival of the pump pulse. The gas mixtures were flowed slowly through the cell to reduce the effects of this, but these pressure waves can play a significant role in vibrational energy exchange processes, thus hindering the performance of indirectly pumped CO$_2$ laser and limiting the maximum total pressure to <4 atm in this experiment. A fast flow of the gas may be able to mitigate this effect, but no extra modifications have been tested in this study.

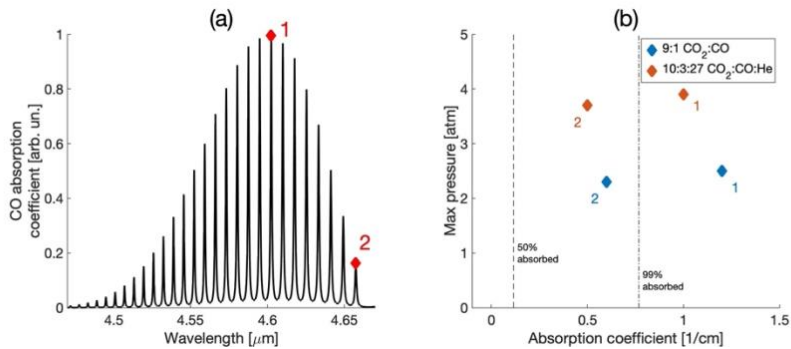

Figure 7. (a) Normalized absorption in 1 atm CO. Red markers indicate pump wavelengths used in experiment. (b) Maximum lasing pressure vs estimated experimental absorption coefficient. Numbers correspond to pump wavelengths indicated in (a).

### 3.4 Possibility of repetition rate scaling

While the measurements above have been conducted at a relatively low repetition rate of 3 Hz, it is important to identify the rate at which the gas molecules in the system recover from the disturbance of pressure waves caused by the absorption and dissipation of heat caused by the energetic pump pulse. Figure 8 indicates the results of a He-Ne laser visible beam deflection experiment used to characterize this relaxation rate in a gas mixture of 0.75 atm $CO_2$ and 9.25 atm He pumped at 4.4 µm.

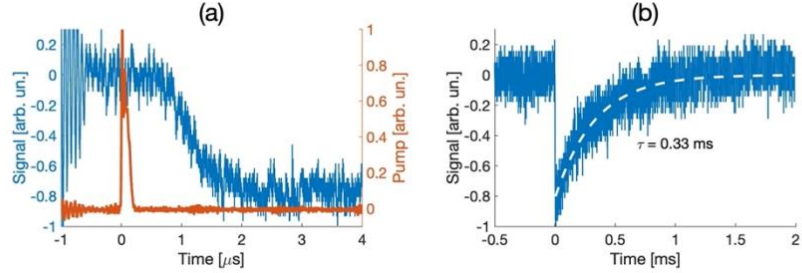

Figure 8. (a) Deflected probe signal (in blue) and Fe:ZnSe pump pulse (in red). (b) Deflected probe signal and recovery. The dashed white line indicates an exponential fit with a time constant of 0.33 ms.

Figure 8(a) shows the probe signal in blue and the Fe:ZnSe laser pump pulse in red on a microsecond time scale. It can be seen that immediately after the ~200 ns pump pulse, the probe signal remains unaffected, but ~1 µs after the pump pulse, the probe beam is deflected, and the signal is reduced. Figure 8(b) shows the recovery of the probe signal for a similar pump pulse energy on a longer millisecond time scale. It can be seen that after ≥1 ms, the gas system has returned to equilibrium and the visible probe signal returns to the initial value. Fitting an exponential curve gives a relaxation time constant of ~0.33 ms at this 10 atm total pressure. A similar result was measured in $CO/CO_2$ mixes, where the relaxation time constant was measured to be ~0.50 ms in a mix of 1 atm CO, 0.1 atm $CO_2$, and 8.9 atm He. Given that this relaxation time is inversely proportional to pressure, it can be estimated that for the high pressures (>10 atm) desirable for short-pulse amplification, the system will reset in ≤1 ms. This indicates that an optically pumped $CO_2$ laser system is capable of operating at a repetition rate of ~1 kHz.

There are currently two feasible options for the development of a 1 kHz solid-state source capable of pumping a $CO_2$ active medium for the amplification of 10 µm pulses to GW level powers (with a few mJ in a few ps). A 1 kHz Fe:ZnSe laser system pumped by Er:YAG lasers providing mJ-level output has already been developed [9], and could be scaled to provide tens of mJs utilizing commercially available 50 mJ, 1 kHz Er:YAG lasers for pumping. Alternatively, the efficient frequency down-conversion of 1064 nm Nd:YAG laser pulses could be used to generate pulses at 4.256 µm, the fourth sub-harmonic of this wavelength. Nd:YAG lasers are capable of operating at a kilohertz repetition rate, and efficient conversion of 15 ns pulses from 1064 nm to 2128 nm and from 2.1 µm to ~3-5 µm has already been demonstrated with optical parametric oscillators operating at or near the degeneracy point [12,13]. It should be noted that for such a $CO_2$ laser system producing multi-watt average powers, a certain heat exchanger similar to that used in industrial $CO_2$ pulsed lasers should be designed and utilized.

## 4. Conclusions

In this paper, lasing is studied in a $CO_2$ active medium optically pumped with a tunable Fe:ZnSe laser at 3 Hz. It is experimentally demonstrated that by tuning the pump wavelength to reduce the absorption coefficient, the maximum pressure at which lasing is observed can be increased to 15 atm. The optimum pump wavelength is found to be 4.40 µm, which conveniently allows for propagation of the pump pulse in air without loss of energy due to negligibly small $CO_2$ absorption. A high optical-to-optical conversion efficiency of ~10% is measured at 7 atm, falling to ~5% at pressures above 10 atm. This efficiency could be further optimized using a two-pass or transverse pumping scheme, as only ~70% of the total pump energy was absorbed at optimal conditions. Gain lifetime is measured as a function of pressure, and the ~1 µs lifetime observed at high pressures indicate the feasibility of using optically pumped $CO_2$ as a multi-pass or regenerative amplifier. With the development of more energetic pump sources at ~4.4 µm and future work performed to optimize the design, such an amplifier may be used to achieve TW level power in a picosecond 10 µm pulse. An alternative pumping scheme in which $CO_2$ is indirectly excited via pumping a collisional partner molecule CO is also studied, but it is determined that low gain coefficients and the generation of heat or pressure waves within the cell hinder the performance of this method. Finally, He-Ne probe beam deflection measurements demonstrate that a high-pressure $CO_2$ laser can be pumped at a maximum repetition rate of ~1 kHz. This relatively fast repetition rate and the observation of significant gain and good efficiency at and above 10 atm make a high-pressure optically pumped $CO_2$ medium a promising candidate for short pulse amplification [18].

## Funding

This material is based on work supported by the Office of Naval Research (ONR) MURI (No. N00014-17-1-2705) and the Department of Energy (DOE) Office of Science Accelerator Stewardship Award No. DE-SC0018378.

## Disclosures

The authors declare no conflicts of interest.

## Data availability

The data that support the findings of this study are available from the corresponding author upon reasonable request.